\documentclass[runningheads,citeauthoryear]{apinv1}
\usepackage{epsfig,cite,graphics}
\usepackage{marvosym} 
\usepackage[utf8]{inputenc}
\usepackage{color}

\begin{document}

\title{Interstellar extinction toward symbiotic stars }
\titlerunning{Symbiotic stars}
\author{R. K. Zamanov, V. D. Marchev, K. A. Stoyanov}
\authorrunning{Zamanov, Marchev, Stoyanov}
\tocauthor{R. K. Zamanov\inst{1}, V. D. Marchev\inst{1}, K. A. Stoyanov\inst{1}} 
\institute{Institute of Astronomy and NAO, Bulgarian Academy of Sciences, BG-1784, Sofia   \newline
	\email{rkz@astro.bas.bg}    }
\papertype{Research report, Submitted on 21 May 2020; Accepted on 25 June 2020}	
\maketitle

\begin{abstract}
  Using diffuse interstellar bands (DIBs) at 5780~\AA, 5797~\AA\ and 6613~\AA, visible in the high 
  resolution spectra, and measuring their equivalent widths,
  we estimate the interstellar extinction toward seven symbiotic stars.
  We find  $E_{B-V}$= $1.28 \pm 0.10$ for AS~289, 
           $E_{B-V}$= $1.55 \pm 0.10$ for BI~Cru,
           $E_{B-V}$= $0.63 \pm 0.10$ for HD~330036,
	   $E_{B-V}$= $0.33 \pm 0.05$ for V2756~Sgr,
	   $E_{B-V}$= $0.30 \pm 0.05$ for V2905~Sgr, 
	   $E_{B-V}$= $1.52 \pm 0.11$ for V417~Cen,
	   $E_{B-V}$= $0.81 \pm 0.10$ for PN~Sa~3-22.
  The derived values are in agreement with the extinction through the Galaxy. 
\end{abstract}

\section{Introduction}
The symbiotic stars are interacting binaries, in which an evolved  red-yellow giant or supergiant
transfers mass to hot compact companion, usually an white dwarf (Miko{\l}ajewska 2012; Akras et al. 2019).
The cool primary can be a normal red giant, yellow giant or a Mira-type asymptotic giant branch star.
The two components are embedded in complex surroundings, 
such as ionized and neutral regions of the wind of the giant, 
accretion disc, in some cases colliding winds, non-relativistic jets, remnant from nova explosion, dust shell
(Sokoloski et al. 2017). 

A few diffuse interstellar bands (DIBs) are clearly visible in our spectra of symbiotic stars. 
Diffuse interstellar bands are over 400 broad spectroscopic absorption features 
observed in stellar spectra in ultraviolet, visible and infra-red ranges (Sarre 2006, Fan et al. 2019). 
DIB carriers are  believed to be large molecules in gas phase
(e.g. Spieler et al. 2017;  Elyajouri et al. 2018).

Using  high resolution optical spectra of a few symbiotic stars
we estimate the interstellar extinction toward them.

\section{ Observations}

The observations have been performed  with FEROS at the 2.2m 
telescope  La Silla,  European Southern  Observatory (ESO).   
FEROS is a fibre-fed echelle spectrograph, providing a high resolution of 
$\lambda/\Delta \lambda=$48000, 
a wide wavelength coverage from about 4000~\AA\  to 8900~\AA\  in one exposure 
and a high throughput (Kaufer et al. 1999). 
The 39 orders of the echelle spectrum are registered with a 2k$\times$4k EEV CCD. 
The spectra are reduced using the dedicated FEROS data reduction software 
implemented in the ESO-MIDAS system. 
A few examples of the interstellar features in our spectra are given on 
Fig.~\ref{f.BICru}, Fig.~\ref{f.HD33}, and  Fig.~\ref{f.AS289}.
The most prominent interstellar features are  NaD1 and D2 lines. 
In Fig.~\ref{f.BICru} the NaD1 and D2 lines are plotted in the upper panel 
together with KI7699 line. In the lower panel are plotted the DIB5780
and DIB6613. The comparison between the profiles indicates that 
a part of the NaD lines has circumstellar origin (not interstellar). 
Thus for estimation of the interstellar extinction we will use the DIBs.

\section{Interstellar reddening  $E_{B-V}$}

There is an apparent general correlation between the DIB strengths and
the interstellar extinction or reddening.  
To calculate $E_{B-V}$ we measure the equivalent width of a few interstellar features
and use the following equations 
from  Puspitarini, Lallement \& Chen (2013):

\begin{equation}
  E_{B-V}  =  2.3 W_{5780} + 0.0086,                  
\end{equation}

\begin{equation}
  E_{B-V}  = 6.3 W_{5797} + 0.0203,          
\end{equation}

\begin{equation}
  E_{B-V}  = 5.1 W_{6613} + 0.0008.   
\end{equation}
These equations refer to the equivalent width of   
DIB~$\lambda$5780.38, DIB~$\lambda$5797.06, 
DIB~$\lambda$6613.62, 
respectively. In them W is in \AA\  and $E_{B-V}$ is in magnitudes. 
The DIB 5797 is more tightly correlated with column density of molecular hydrogen
while the DIB 5780 -- with that of atomic hydrogen (Weselak 2019).
The DIB 6613 total column density is proportional
to the total column density of Ca II and H~I (Sonnentrucker et al. 1999). 

For each object we measured equivalent widths of the three DIBs. 
We also approximated the red giant features in and around DIBs with 
stellar spectra of similar spectral types taken from the 
ELODIE archive (Moultaka et al. 2004) 
and field stars across the  Hertzsprung -- Russell  diagram (Bagnulo et al. 2003). 

\begin{figure}    
   \vspace{7.0cm}     
   \includegraphics{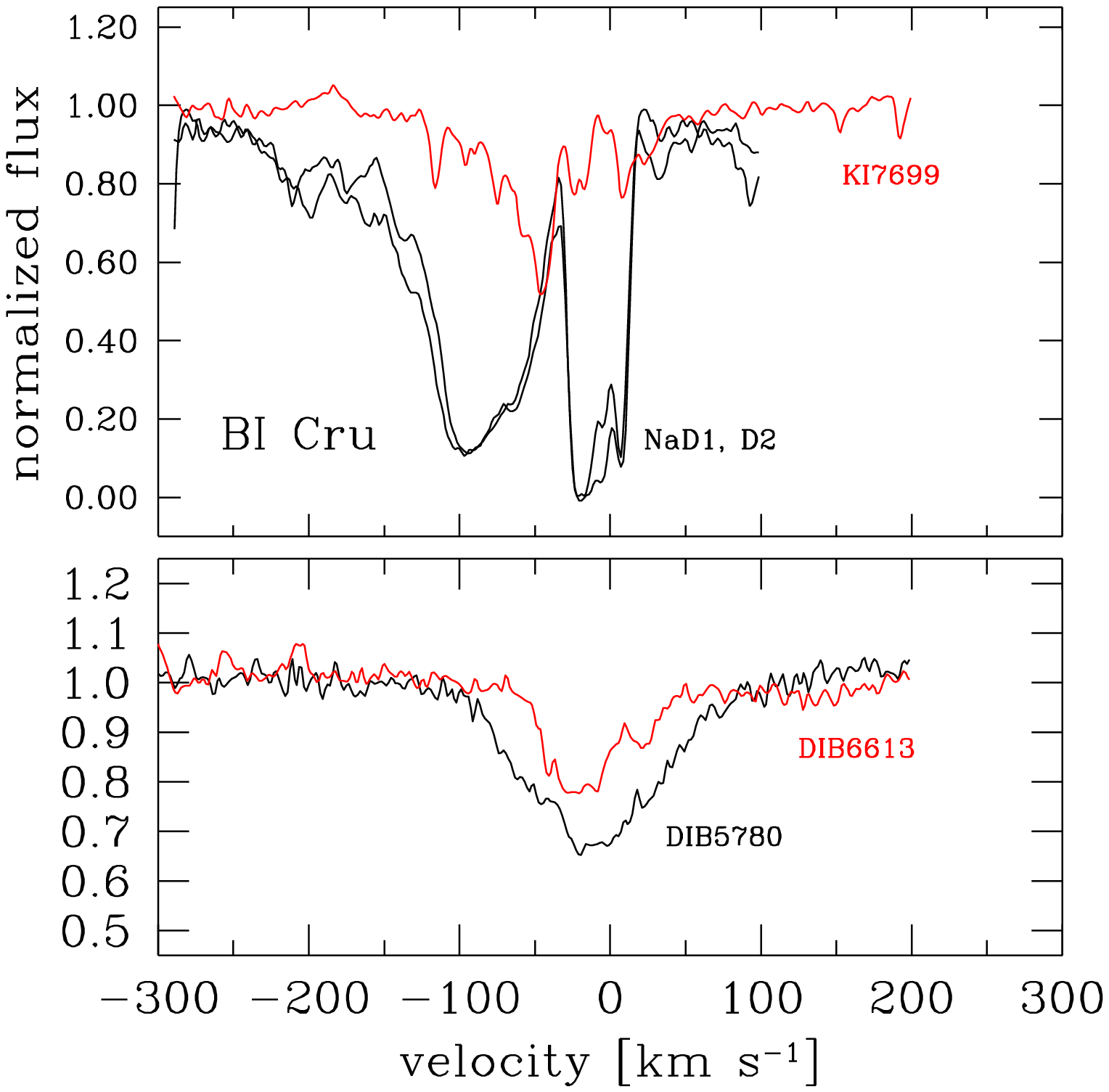} 	  
   \caption[]{Interstellar lines in the spectrum of BI~Cru - NaD1 and D2 lines, KI7699, DIB 6613, DIB 5780. 
   }
   \label{f.BICru}
   \vspace{6.0cm}
   \includegraphics{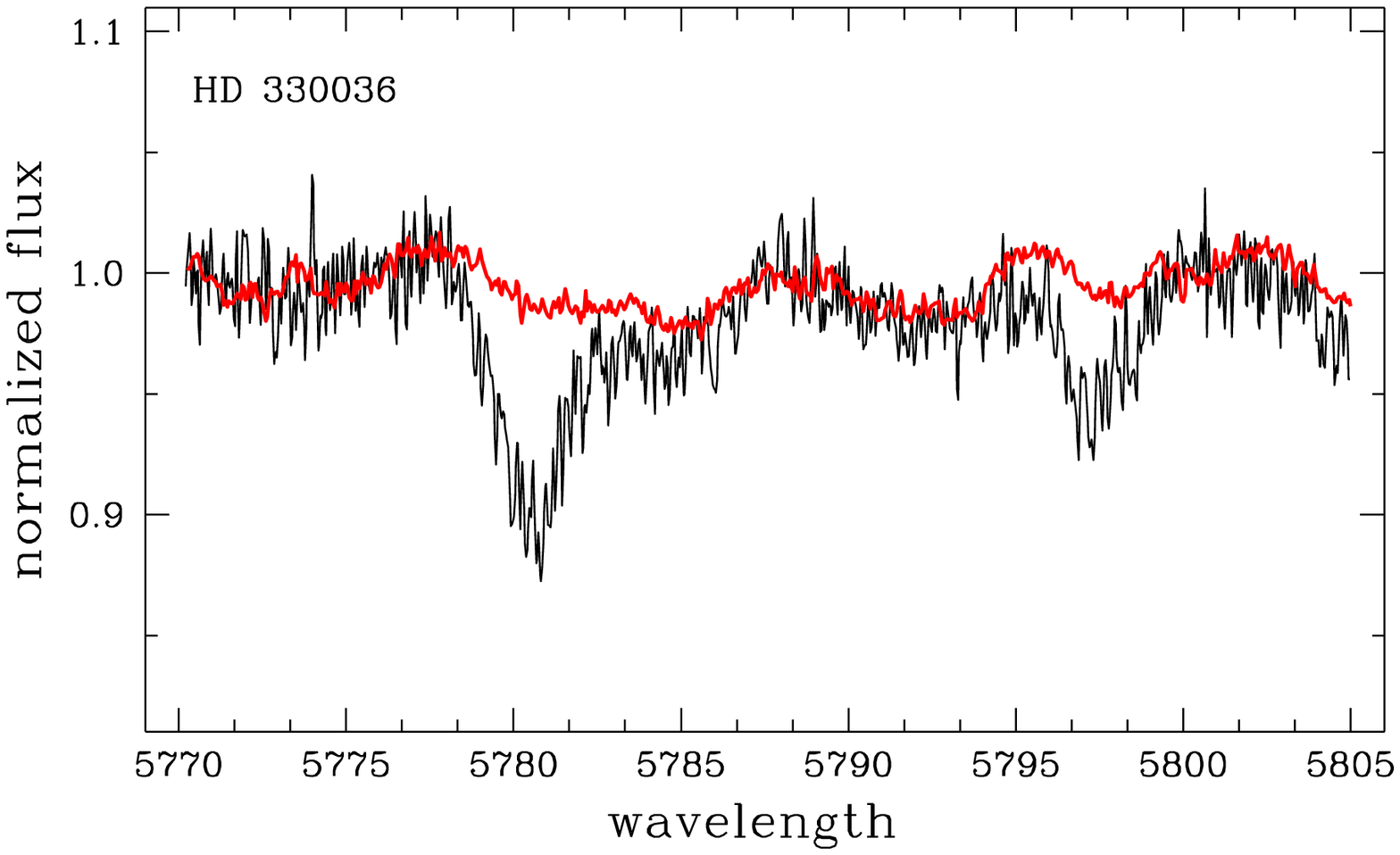}	   
   \caption[]{Spectrum of HD~330036 (black solid line) compared with HD~432 (the spectrum is broadened, red  line). 
   The comparison gives the possibility 
   to measure precisely the equivalent widths of DIB5780 and DIB5797.   }
   \label{f.HD33}	   
 \end{figure}

In Fig.~\ref{f.HD33}, the spectrum of HD~330036 (black solid line)
is  compared with the bright giant HD~432 ($\beta$~Cas,  F2III). The spectrum 
of $\beta$~Cas is broadened to take into account the fast rotation 
of the cool component of HD~330036. 
This figure demonstrates how the comparison with field giant gives the possibility 
to  place the continuum, and to measure more precisely the equivalent widths of DIB5780 and DIB5797, which 
are used in Eq.1 and Eq.2.

\begin{figure}    
   \vspace{7.5cm}
   \includegraphics{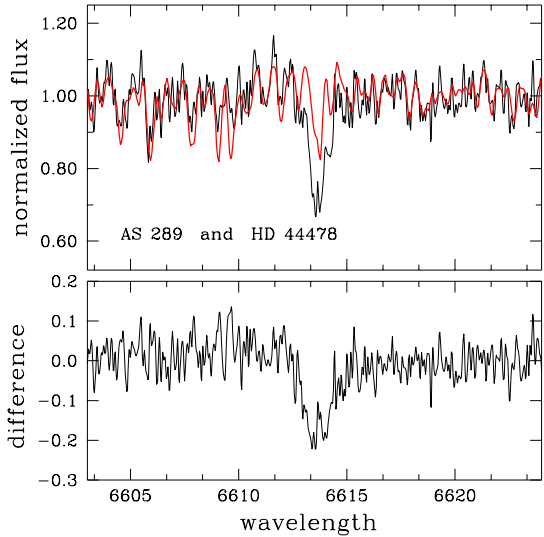}	    
   \caption[]{Spectrum of AS~289 (black solid line) compared with HD~44478 
   (red line). The lower panel is the difference between them. 
   }
   \label{f.AS289}	
  \end{figure}

In  Fig.~\ref{f.AS289} the spectrum of AS~289 is plotted together with 
that of the bright red giant HD~44478 ($\mu$~Gem, M3IIIab). 
This figure illustrates that the subtraction of the red giant spectrum 
gives us the possibility to estimate accurately the equivalent width of the DIB~6613. 

In Table~1 are given the name of the object, 
the modified Julian Day (MJD) of the observation as given in the fits file header, 
and the spectral type of the cool component.   
The last column is the upper limit of  $E_{B-V}$,   
which is the extinction through our Galaxy in the direction of the object, 
taken from the IRSA: Galactic Reddening and Extinction Calculator 
in the NASA/IPAC Extragalactic Database (NED),  
which  is operated by the Jet Propulsion Laboratory,  California Institute of Technology. 
This calculator uses Galactic reddening maps to determine the total Galactic line-of-sight reddening
and is based on the results by  Schlegel, Finkbeiner \& Davis (1998)  
and Schlafly \& Finkbeiner (2011). 

Our measurements are summarized in Table~2. 
In this table are given name of the object,
$W_{5780}$, $W_{5797}$, $W_{6613}$ and $E_{B-V}$ calculated for each DIB.
The last column is the average value of $E_{B-V}$.



\setlength{\tabcolsep}{7pt}

\begin{table*}
\caption{In the table are given name of the object, MJD of observations, spectral type of the cool component,   
the upper limit of the reddening (IRSA values, see Sect.~3).  }
\begin{tabular}{c | c  c  |   c c  c c}
\\
No:    &  object       &  MJD	    &  spec.type  & upper limit      &  \\		  
       &     	       &	    &		  &  of $E_{B-V}$       &  \\				  
\hline
       &     	       &	    &		     &  	     &  \\	  
1      &  AS 289       & 53161.3071 &  M3.5~III  &  2.44 -  2.84     &  \\   
2      &  BI Cru       & 53107.2002 &  M2/3~III  &  5.78 -  6.72     &  \\   
3      &  HD 330036    & 53107.2882 &  F5~III	 &  1.06 -  1.23     &  \\    
4      &  V2756 Sgr    & 53164.3968 &  M4~III	 &  0.27 -  0.31     &  \\    
5      &  V2905 Sgr    & 53184.0858 &  M3~III	 &  0.27 -  0.32     &  \\   
6      &  V417 Cen     & 53107.2548 &  F6V-G9Ib  &  7.63 -  8.87     &  \\    
7      &  PN Sa 3-22   & 53402.2405 &  M4.5~III  &  0.81 -  0.94     &  \\   
&      &     	       &	    &		 &		     &  \\ 
\hline
\end{tabular}
\end{table*}

\begin{table*}
\caption{Equivalent widths of the DIBs [\AA]  and calculated $E_{B-V}$ [magnitudes] of the symbiotic stars. 
In the table are given name of the object, EWs of the DIBs and the calculated $E_{B-V}$. 
 The last column are our estimations of $E_{B-V}$, i.e.  the mean values. 
 }
\begin{tabular}{ c  c  | c  |  c  |  c  |  c c | }
\\
\hline 
      &              &                      &                       &                      &  & 		 \\
 No:  &  object      & $W_{5780}$ $E_{B-V}$ & $W_{5797}$  $E_{B-V}$ & $W_{6613}$ $E_{B-V}$ &  &   $E_{B-V}$	 \\ 
      &     	     & DIB~$\lambda$5780    & DIB~$\lambda$5797     &   DIB~$\lambda$6613  &  &     mean	  \\ 
      &              & [\AA] [mag]          & [\AA] [mag]           & [\AA] [mag]          &  &    [mag]	 \\
      &              &                      &                       &                      &  & 		 \\  
\hline		    
      &              &                      &                       &                      &  & 		 \\	     
1     &  AS 289      &  0.576  1.33	    &   ---	            & 0.240  1.22          &  & $1.28\pm0.07$	 \\
2     &  BI Cru      &  0.697  1.61	    & 0.250  1.60           & 0.288  1.47          &  & $1.55\pm0.10$	 \\  
3     &  HD 330036   &  0.265  0.62	    & 0.109  0.71           & 0.110  0.56          &  & $0.63\pm0.10$	 \\ 
4     &  V2756 Sgr   &  0.141  0.33	    &    ---	            &    ---               &  & $0.33\pm0.05$	 \\ 
5     &  V2905 Sgr   &  0.146  0.34	    &    ---	            & 0.050  0.26          &  & $0.30\pm0.05$	 \\ 
6     &  V417 Cen    &  0.674  1.56	    & 0.263  1.68           & 0.272  1.39          &  & $1.52\pm0.10$	 \\
7     &  PN Sa 3-22  &  0.360  0.84	    & 0.125  0.81           & 0.154  0.79          &  & $0.81\pm0.10$	 \\
      &              &                      &                       &                      &  & 		 \\	     
\hline
\end{tabular}
\end{table*}

\section{Objects and results}
As a reference point and upper limit of the  expected reddening we use  $E_{B-V}$
through the Milky Way in the direction of the objects. 
Because IRSA calculates the extinction through the entire  Galaxy, it is
an upper limit for objects in our Galaxy. 
In some cases all the interstellar clouds are in front of the star 
and it will give a similar result to our. 
In other cases (e.g. AS 289, BI Cru and V417 Cen), part of the clouds are behind 
the object and it leads to a significant difference between our estimation and 
IRSA values.

{\bf AS~289 (V343 Ser):} for this object  Luna \& Costa (2005) give $E_{B-V}$=1.18.
Our value   $E_{B-V}=1.28 \pm 0.07$  is similar. 

{\bf BI Cru (Hen 3-782):}  for this object Rossi et al. (1988)  give   $ E_{B-V} =1.5 \pm 0.5$.
Our value $E_{B-V} =1.6 \pm 0.1$ is similar and with better accuracy. 

{\bf HD330036 (PN Cn 1-1)}  having three circumbinary dust shells 
is a demonstration of the  complexity of the symbiotic environment
(Angeloni  et al. 2007).    
For this object  Lutz (1984) estimated $E_{B-V}= 0.28$, 
Bhatt \& Mallik (1986) gave  $E_{B-V}=0.41$. 
Our value  $E_{B-V} = 0.63 \pm 0.10$ is higher, but still consistent with the extinction through the Milky Way. 

{\bf V2756~Sgr (AS 293):} for this object we can measure only DIB5780. The derived extinction $E_{B-V} = 0.33\pm0.05$	 is 
consistent with the Galactic reddening and with the value $E_{B-V} = 0.32 $
estimated by Luna \& Costa (2005). 

{\bf V2905~Sgr (AS 299):} for this object  Pereira, Landaberry \& da Conceicao (1998) measured  $ E_{B-V} = 0.72 \pm 0.15$, 
Luna \& Costa (2005) --  $E_{B-V} =0.43$.
We compared the spectrum of V2905~Sgr, with that of  the red giant 
51 Gem, for which SIMBAD gives  M4III, a spectral class similar to that of V2905~Sgr.
DIB~5780 and DIB6613 are clearly detectable and correspond to $E_{B-V} = 0.30 \pm 0.05$.
We do not detect DIB 5797, its EW is expected to be low ($EW (5797) \le 0.03$~\AA). 
The reddening through the Galaxy in the direction of V2905~Sgr is $E_{B-V} = 0.27-0.32$, 
which is in agreement with our value $E_{B-V} = 0.30\pm0.05$.  

{\bf V417~Cen (Hen 3-977): } for this object 
Stoyanov et al. (2014) estimated  $E_{B-V}$= $0.95 \pm 0.10$ on base of the non-variable part of KI7699 line.
Here using three DIBs, we obtain a larger and probably more accurate
value  $E_{B-V}$= $1.52\pm0.10$. 

{\bf PN Sa 3-22} 
(also named in few papers  {\bf St 2-22})  is a symbiotic star with high-velocity ($\approx\!1700$~km~s$^{-1}$)
bipolar jets (Tomov et al. 2017). 
Mikolajewska, Acker \& Stenholm (1997)  have derived  reddening $E_{B-V}$=1.0 mag. 
Our value $E_{B-V}$=$0.81 \pm 0.1$ is slightly lower
and more consistent with the Galactic reddening.

\section{Conclusion}
We measured the interstellar extinction toward 7 symbiotic stars using the equivalent widths of DIBs
visible in our FEROS spectra. 
Our values are in agreement with the extinction 
through the Galaxy and are improvement over the previous measurements. 
They should be useful for modeling.
\\
\vskip 0.3cm 

{\bf Acknowledgments: } 
This work is part of the project  K$\Pi$-06-H28/2 08.12.2018
"Binary stars with compact object" supported by National Science Fund (Bulgaria).
It is based on observations 
obtained in ESO programmes 073.D-0724A and 074.D-0114.
We thank to the anonymous referee for the useful comments.

\end{document}